\documentclass[twocolumn,runningheads]{astronepal}
\smartqed
\usepackage{graphicx}
\usepackage{amsmath}
\usepackage{amsfonts}
\usepackage{amssymb}

\def\eg{{\it e.g.,} }
\def\etal{{\em et al.} }
\def\ie{{\em i.e.,} }
\def\lsim{\lower.5ex\hbox{$\; \buildrel < \over \sim \;$}}
\def\gsim{\lower.5ex\hbox{$\; \buildrel > \over \sim \;$}}

\def\ee{{$e^--e^+$}}
\def\ep{{$e^--p^+$}}
\begin{document}

\title{How Plasma Composition affects the relativistic flows and the emergent spectra } 

\titlerunning{Plasma Composition}        

\author{Indranil Chattopadhyay, Sudip Garain, Himadri Ghosh}

\institute{I. Chattopadhyay,\at
              ARIES, Manora Peak, Nainital 263129, India. \\ 
              \email{indra@aries.res.in}         
\and          S. Garain, H. Ghosh \at
              S. N. Bose National Center for Basic Sciences\\
              \email{sudip@bose.res.in, himadri@bose.res.in}}

\maketitle

\begin{abstract}
It has been recently shown that transonic electron positron
fluid is the least relativistic, compared to the fluid containing finite proportion of baryons.
We compute spectra from these
flows in general relativity (GR) including the effect of light bending.
We consider the bremsstrahlung process to supply the seed photons.
We choose accretion in the advective domain, and for simplicity the radial accretion
or Bondi type accretion. We show that electron positron accreting
flow produces the softest spectra and the lowest luminosity.
\end{abstract}


\section{Introduction}

Relativistic flows are likely to be encountered in accretion discs around compact objects like
neutron stars and black holes, astrophysical jets, Gamma Ray Bursts (GRBs) etc. A fluid is said to be relativistic
if the bulk speed of the plasma is comparable to the speed of light ($c$), or if its thermal energy is comparable or
greater than its rest energy --- a fancy way of saying, that the random or jittery speed of the constitute
particles of the fluid become comparable to the speed of light. This brings the issue of equation of state (EoS)
of the fluid.
The issue of relativistic equation of state was raised long ago (Chandrasekhar 1938, hereafter C38;
Taub 1948; Synge 1957, hereafter S57),
and was used for theoretical
calculations too (Blumenthal \& Mathews 1976, hereafter BM76; Fukue 1987), although didnot become very popular in the
community.
Later Falle \& Komissarov (1996) showed that the relativistic
EoS of the form presented by C38 and S57, is computationally expensive which led 
Mignone \etal (2005) to use
the EoS of the form presented by BM76. Ryu \etal (2006, hereafter RCC06), proposed a new EoS, which is very close to the EoS by
C38 and S57,
and better than the EoS proposed by BM76, and which can be efficiently implemented in a simulation code.
However, these EoS states are for single species fluid. Chattopadhyay (2008; hereafter C08), Chattopadhyay \& Ryu (2009;
hereafter CR09) then modified the single species EoS for multi-species EoS, and analytically showed that the 
the accreting and solutions around black holes are indeed dependent on the composition of the flow. More interestingly,
CR09 showed that electron-positron jets are the least relativistic when compared with flows containing protons. The most
relativistic flow is the one with composition parameter (defined as the ratio proton to electron number density) $\xi\sim0.24$.
Later Chattopadhyay \& Chakrabarti (2011; hereafter CC11) showed that, fluids containing protons can produce accretion shocks while
an entirely leptonic fluid ($\xi=0$) will not be able to form accretion shocks. Although it is difficult to envisage 
an accreting flow entirely composed of leptons \ie electrons-positrons from close to the horizon to infinity,
however, our main intention is to show that --- (1) fluid behaviour do depend on composition even in absence of
composition, and (2) it is erroneous to a assume the most relativistic matter fluid is the lightest one that is
pair plasma. In this paper therefore we want to compute the radiation produced by flow accreting onto a black hole,
and show that indeed the spectra also depends on the composition of the fluid even if the outer boundary condition is the same.
For that purpose we have developed a general relativistic Monte-Carlo code.

%
%
\section{Governing Equations}

We assume a non-rotating black hole described by the Schwarzschild radii
\begin{eqnarray}\label{}
ds^2 & = & -\left(1-\frac{2GM_B}{c^2r} \right)c^2dt^2 \\ \nonumber
&+ & \left(1-\frac{2GM_B}{c^2r}\right)^{-1}dr^2
+r^2d{\theta}^2+r^2\sin^2{\theta}d\phi^2,
\end{eqnarray}
where, $G$, $M_B$ and $c$ are the universal gravitational constant, the mass of the central black hole
and the speed of light, respectively, and $t,~r,~\theta,~\phi$ are the usual four coordinates.
\begin{equation}\label{}
T^{\mu \nu}_{; \nu}=-T^{\mu \nu}_R ~~~~ \mbox{and} ~~~~ (nu^{\nu})_{; \nu}=0,
\end{equation}
where, the energy momentum tensor of the fluid is $T^{\mu \nu}=(e+p)u^{\mu}u^{\nu}+pg^{\mu \nu}$,
the radiation stress-tensor being $T^{\mu \nu}_R$, where $N$ is the total number density of the fluid.
The fluid equation is solved theoretically assuming there is no radiation. And then it is fed to the general
relativistic Monte Carlo code. The Monte Carlo code estimates bremsstrahlung emission from each grid point, and then
the emitted photons travels and impinges on the electron at some other point. If the energy of the photon is less
than electrons kinetic or thermal energy then inverse-Comptonization, \ie the photon will take energy away.
If the the photon energy is more, then the opposite happens. The energy density of the fluid is given by (see,
C08, C09), 
\begin{eqnarray}
 e &=& n_{e^-}m_ec^2f \\
f &=& (2-\xi)\left[1+\Theta\left(\frac{9\Theta+3}{3\Theta+2}\right)\right] \\ \nonumber
&+& \xi\left[\frac{1}{\eta}+\Theta\left(\frac{9\Theta+3/\eta}{3\Theta+2/\eta}
\right)\right],
\end{eqnarray}
where, $\xi=n_{p^+}/n_{e^-}$ is the composition parameter, where electron number density $n_{e^-}=0.5$, the mass density
 $\rho=n_{e^-}m_e\left\{ 2-\xi
\left(1-1/\eta \right)\right\}$, $\eta=m_e/m_{p^+}$, $\Theta=kT/m_ec^2$ and the pressure $p=2n_{e^-}kT$. For radial and
adiabatic flow in steady state, the equation of motion simplifies to the form given in CR09. 

The initial configuration of the fluid is, semi-analytical calculation of an adiabatic radial inflow
for Bernoulli parameter ${\cal E}=0.001$ in units of $c^2$. The the electron number
density is plotted assuming $M_B=10M_{\odot}$
and particle flux rate to be ${\dot N}=1.724\times 10^{42}$cm$^{-3}$s$^{-1}$. This value ${\dot N}$ is
the particle flux corresponding to accretion rate $0.1{\dot M}_{\rm Edd}$ for an electron-proton (\ep) fluid or
a fluid with $\xi=1$. All types of fluid starts with these values of ${\dot N}$ and ${\cal E}$.
In Fig. 1a-b, we plot total lepton density $(n_{e^-}-n_{e^+})$ (long dashed), temperature $T$ (dashed)
and the radial three velocity $v$ (solid) are plotted with $r$ in log-log scale, for two kinds of fluid
the $\xi=0$ or \ee (a) and $\xi=1$ or \ep (b). It is to be noted the positron number density ($n_{e^+}$)
is different for these two fluid, \eg for $\xi=0$, $n_{e^+}=n_{e^-}$ and $n_{p^+}=0$ but for 
$\xi=1$, $n_{e^+}=0$ and $n_{p^+}=n_{e^-}$.
\begin{figure}[h!]
\vspace{0.0cm} \centering
\includegraphics[height=4.0cm]{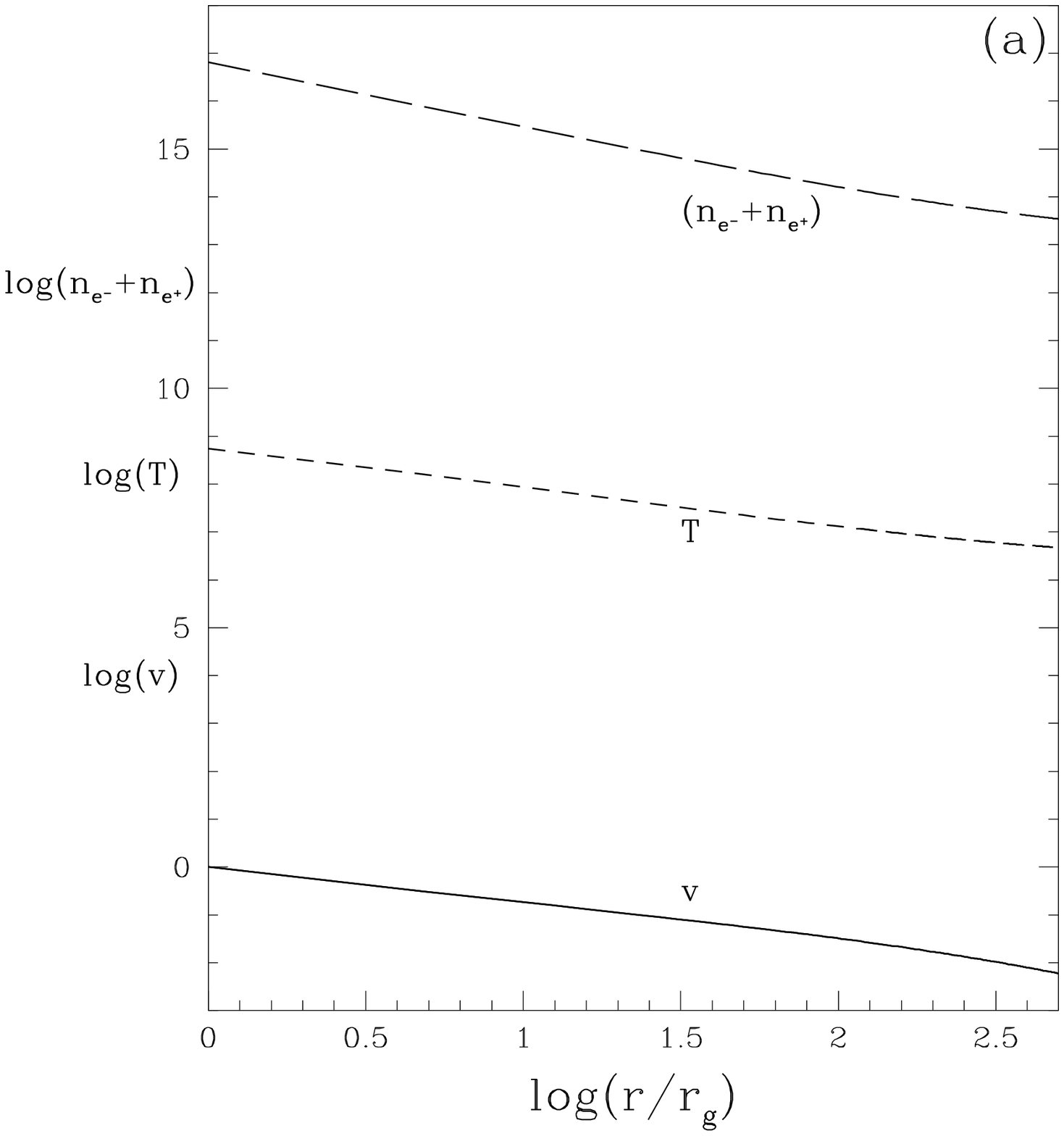}
\includegraphics[height=4.0cm]{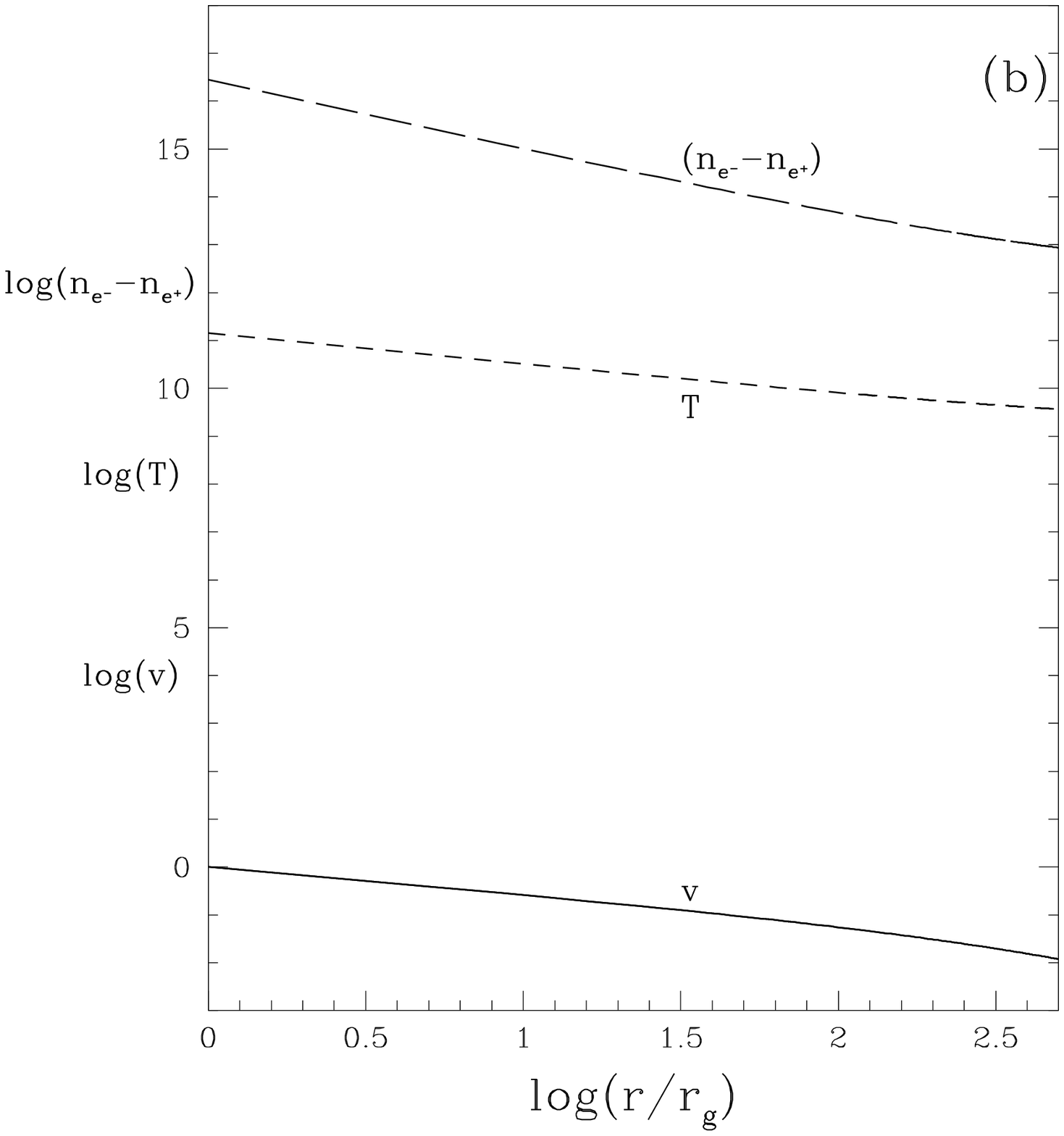}
\caption{ Total lepton number density $(n_{e^-}-n_{e^+})$ (long dashed), temperature $T$ (dashed)
and the radial three velocity $v$ (solid) are plotted with $r$ in log-log scale, for two kinds of fluid
the $\xi=0$ or \ee (a) and $\xi=1$ or \ep (b). Both the fluids starts with ${\dot N}=1.724\times 10^{42}$cm$^{-3}$s$^{-1}$
and ${\cal E}=0.001$ in units of $c^2$.}
\end{figure}

Starting with these initial values of the fluid, we calculate the Bremsstrahlung seed photon distribution.
The spectral form of the bremsstrahlung emission from $e^{\pm}-p^+$, $e^{\pm}$ and \ee are (Gould 1980)
\begin{eqnarray}
\left(\frac{dE}{dVdtd\epsilon}\right)_{e^{\pm}-p^+}&=&\frac{16}{3}{\sqrt{\frac{2}{\pi}}}
\alpha^3\lambda^2c(2-\xi)\xi n^2_{e^-}\Theta^{-1/2} \\ \nonumber
& & e^{-a}K_0(a)\left[1+\frac{\epsilon}{4m_ec^2}f(a) \right], \\
\left(\frac{dE}{dVdtd\epsilon}\right)_{e^{\pm}}&=&\frac{128}{15}\alpha^3\lambda^2[1+(1-\xi)^2]n^2_{e^-} \\ \nonumber
&& {\sqrt{\frac{kT}{m_e\pi}}}ae^{-a}K_0(a)\Psi(a), \\
\left(\frac{dE}{dVdtd\epsilon}\right)_{e^--e^+}&=&\frac{128}{15}\alpha^3\lambda^2(1-\xi)n^2_{e^-} \\ \nonumber
&& {\sqrt{\frac{kT}{m_e\pi}}}ae^{-a}K_0(a)\Psi(a),
\end{eqnarray}
where, $a=\epsilon/2kT$, $\epsilon=h\nu$,
$\alpha=e^2/\hbar c$, $\lambda=\hbar/m_ec$, and
$f(a)=1/4a+3K_1(a)/K_0(a)+a[1-K_2(a)/K_0(a)]$,
$\Psi(a)=3/4a+K_1(a)/K_0(a)+c_2e^{-a}/16K_0(a)$. The $K$'s are modified Bessel's function of various kind.
Now there are at least three reference frames involved, the local comoving frame, the local rest frame and the Schwarzschild
frame. The three velocities are measured in the local fixed frame and the cooling rates are calculated in the comoving
frame. The photons, however will be moving in a curved geometry, and would follow geodesic equations. 
The coordinate transformations between the locally fixed frame ($x^{\hat \beta}$) and the coordinate frame
($x^{\mu}$) are given by (Park 2006), 
\begin{eqnarray*}
\frac{\partial}{\partial {\hat t}}& =& \frac{1}{(1-2/r)^{1/2}}\frac{\partial}{\partial  t} \\
\frac{\partial}{\partial {\hat r}}& =& (1-2/r)^{1/2}\frac{\partial}{\partial r} \\
\frac{\partial}{\partial {\hat \theta}}& =& \frac{1}{r}\frac{\partial}{\partial \theta} \\
\frac{\partial}{\partial {\hat \phi}}& =& \frac{1}{rsin\theta}\frac{\partial}{\partial \phi}
\end{eqnarray*}
And the transformation between local fixed frame ($x^{\hat \beta}$) and the comoving frame
($x^{\hat \alpha}_{co}$) are related by
Lorentz transformation of the form
\begin{eqnarray*}
\frac{\partial}{\partial x^{\hat \alpha}_{co}}=\Lambda^{\hat \beta}_{\hat \alpha}(v)\frac{\partial}{\partial x^{\hat \beta}},
\end{eqnarray*}
Where ${\hat \alpha},~{\hat \beta}$ indicates the tetrads and $\Lambda^{\hat \beta}_{\hat \alpha}(v)$
are the components of Lorentz transformation.
The photons in strong gravity moves in curved trajectories and the geodesic equations given by  Weinberg (1972),
can be written as,
\begin{eqnarray}
&& \frac{d^2r}{dt^2}-\frac{3}{2}\frac{1}{r(r-1)}\left(\frac{dr}{dt}\right)^2-(r-1)\left(\frac{d\theta}{dt}\right)^2
\\ \nonumber
&& -(r-1)sin^2\theta \left(\frac{d\phi}{dt}\right)^2+\frac{1}{2r(r-1)}=0 \\
&& \frac{d^2\theta}{dt^2}+\frac{2r-3}{r(r-1)}\frac{dr}{dt}\frac{d\theta}{dt}-sin\theta cos\theta
\left(\frac{d\phi}{dt}\right)^2
=0 \\
&& \frac{d^2\phi}{dt^2}+\frac{2r-3}{r(r-1)}\frac{dr}{dt}\frac{d\phi}{dt}+2 cot\theta \frac{d\theta}{dt} \frac{d\phi}{dt}=0
\end{eqnarray}
Moreover the definition of optical depth in curved space and moving media in radial direction, is modified
to
\begin{eqnarray}
 d\tau=\sigma\gamma \{(2-\xi)n_{e^-}\}(1-vn_r){\sqrt{(1-1/r)}}dt,
\end{eqnarray}
where, $\gamma$, $\sigma$ and $n_r$ are the Lorentz factor, Klein Nishina cross-section and the directional cosine of
the photon.
Equations 5-11, are implemented in the Monte-Carlo code of Ghost \etal (2011) and Garain \etal
(2012), to make the code general relativistic.
Therefore, not only the fluid is moving in curved space, but even the photons are doing the same.
The total cooling/heating of
fluid is $L=B+C$, where, $L$ is cooling term, $B$ is the total bremsstrahlung loss and $C$
is the Comptonization term, if there is heating then $C$ is negative and if there is cooling then 
the sign is positive. $C$ has been calculated following Pozdnyakov \etal (1983). If $\delta E$ is the energy
radiated from a grid from unit volume ($C$ or $B$ or $L$), then the new energy density will be given by,
$e_{n}=e_{old}-\delta E$, which will result in new temperature because $e_n=n_{e^-}m_ec^2f(\xi,\Theta_{\rm n})$,
\begin{eqnarray}
 a_1\Theta^3_{\rm n}+a_2\Theta^2_{\rm n}+a_3\Theta_{\rm n}+a_4=0,
\end{eqnarray}
where, $a_1=54\eta$, $a_2=9[2(\eta+2)-\eta(2-\xi+\xi/\eta)(e_{n}-\rho c^2)/(\rho c^2)-\xi(1-\eta)]$,
$a_3=6[2-(2-\xi+\xi/\eta)(e_{n}-\rho c^2)(1+\eta)/(\rho c^2)]$, and $a_4=-4(2-\xi+\xi/\eta)(e_{n}-\rho c^2)/(\rho c^2)$. The code updates the new non-dimensional temperature of the fluid $\Theta_{\rm n}$,
until it converges. 

\section{Result and Discussion}

\begin{figure}[h!]
\vspace{0.0cm} \centering
\includegraphics[height=9.5cm,angle=-90]{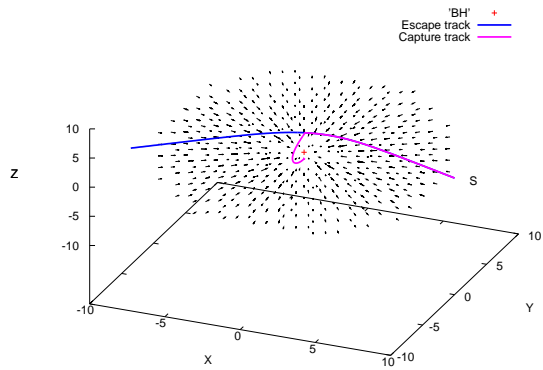}
\caption{Two photons emerging from the outer edge of spherically accreting cloud at S, one escapes
without scattering (blue), the other (magenta) scatters with an electron, changes direction and is absorbed by the
black hole (at $+$). The arrows show the velocity field of the spherical accretion.}
\end{figure}

In Fig. 2,we show that, photon trajectories of two photons originating from same location at S. One
photon escapes and suffers no scattering (blue), while the other scatters (magenta),
changes direction and is captured by the black hole at the centre (location $+$). The arrow heads are the velocity field
or $v$ as shown in Fig. 1a.
In Schwarzschild metric, the trajectories of the photons, which follows geodesic equations, are curved.
It is to be noted that the seed photons generated will depend on $\xi$, as well as temperature and other 
relevant quantities. If the fluid is \ep then $\xi=1$, then the contribution due to Eq. (7) is zero,
while for \ee fluid or when $\xi=0$, contribution due to Eq. (5) is zero. 

\begin{figure}[h!]
\vspace{0.0cm} \centering
\includegraphics[height=8.5cm,angle=-90]{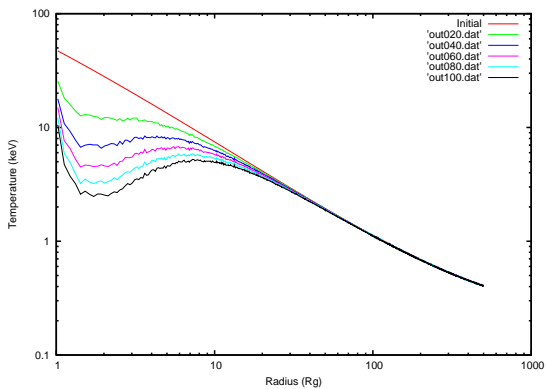}
\caption{Convergence of temperature of \ee fluid or the fluid with $\xi=0$.}
\end{figure}

In Fig. 3, we show the convergence of temperature, or in other words, calculating the new temperature
at each iteration. The initial temperature (red) is decreased as bremsstrahlung and Comptonization is 
considered. At large distances away from the black hole the radiative loss is not much but at $r\lsim 10$
Schwarzschild radii (\ie $r_g=2GM_B/c^2$), cooling becomes important. However, close to the horizon the temperature rises, since the cooling time scale is much larger than
the infall timescale at those distances. The temperature converges after few iterations.
Final spectra is obtained from the converged temperature (black).

\begin{figure}[h!]
\vspace{0.0cm} \centering
\includegraphics[height=8.5cm,angle=-90]{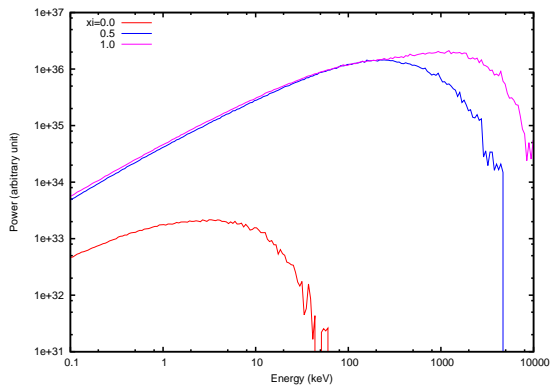}
\caption{Comparison of spectra (power in arbitrary units)
from three different fluids $\xi=0$ (red), $\xi=0.5$ (blue), and
$\xi=1$ (magenta), for same outer boundary condition (see text).}
\end{figure}

In Fig. 4, we present the combined spectra due to bremsstrahlung and Comptonization for fluids
of different composition \eg \ee or $\xi=0$ (red), \ep or $\xi=1$ (magenta), and $\xi=0.5$ (blue).
Clearly, \ee fluid which was coldest and the least relativistic is the least luminous and 
the less energetic in terms of the spectra as well. This is to be expected since the temperature
of \ee fluid is much lower than that by fluids containing protons. Therefore, inverse-Comptonization produces
much less energetic photons. In this paper, we have compared with flows starting with the same
outer boundary condition, namely, at $r_{\rm out}=500~r_g$, ${\cal E}=0.001~c^2$ and ${\dot N}=
1.724\times 10^{42}$cm$^{-3}$s$^{-1}$. We have not injected flows with the same accretion rate,
because in that case the number density of particles at the outer boundary for \ee fluid will be much
higher than fluids with $\xi=0.5$ and \ep fluid.

\section{Conclusion}

It has been shown earlier (C08, CR09, CC11) that, the solution of relativistic, transonic,
and adiabatic accretion depends on the composition of the plasma. More interestingly, we also showed
that, contrary to the expectation, we found that \ee fluid is the least relativistic fluid.
And we conjectured that a purely leptonic flow will be less luminous and of low energy.
Since \ee flow is least relativistic and slowest, density should be higher. This would increase
the optical depth, and therefore should be less luminous. Moreover, the temperature of
\ee is very low too, so higher energies will not
be available through inverse-Comptonization. Chattopadhyay \etal (2013) showed that this is indeed true
for a toy model of radiation, where the seed photons were mono-energetic and artificially injected.
In this paper, we present preliminary solutions of the total spectra from radial
accretion onto black holes by considering realistic seed photons 
(bremsstrahlung) and the Comptonization of those photons, and we vindicate the conclusions of Chattopadhyay \etal (2013).
It may be noted though, that \ee fluid cannot completely describe an accretion flow from infinity to
the horizon. This is shown here as an extreme case, the realistic cases are \ep and $\xi=0.5$
fluid. In view of the results of CC11, extension of these methods to accretion disc, in presence of
all kind of cooling processes, is very interesting. We are working on it and will be reported elsewhere.

\textbf{Acknowledgements} We acknowledge Central Department of
Physics, Tribhuvan University for providing various supports
during the conference.
\\


\begin{thebibliography}{}
\bibitem{bm76} Blumenthal, G. R. \& Mathews, W. G. 1976, ApJ, {\bf 203}, 714 [BM76].
\bibitem{c38} Chandrasekhar, S., 1938,
{\it An Introduction to the Study of Stellar Structure} (NewYork, Dover) [C38].
\bibitem{c08} Chattopadhyay, I. 2008, AIPC, {\bf 1053}, 353. [C08]
\bibitem{cr09} Chattopadhyay, I., Ryu, D., ApJ, 2009, {\bf 694}, 492 [CR09]
\bibitem{cc11} Chattopadhyay, I. Chakrabarti, S. K. 2011, Int. Journ. Mod. Phys. D, {\bf 20}, 1597 [CC11].
\bibitem{cmggkr13} Chattopadhyay, I., Mandal, S., Ghosh, H., Garain, S., Kumar, R., Ryu, D., 2013, Astron. Soc. Ind. Conf. Ser.,
5, 81.
\bibitem{fk96} Falle, S. A. E. G. \& Komissarov, S. S. 1996, MNRAS, {\bf 278}, 586
\bibitem{f87} Fukue, J., 1987, PASJ, 39, 309
\bibitem{ggc12} Garain, S., Ghosh, H., Chakrabarti, S. K., 2012, ApJ, {\bf 758}, 114
\bibitem{gggc11} Ghosh, H., Garain, S., Giri, K., Chakrabarti, S. K., 2011, MNRAS, {\bf 416}, 959
\bibitem{G80} Gould, R. J., 1980, ApJ, {\bf 238}, 1026
\bibitem{mpb05} Mignone, A., Plewa, T., \& Bodo, G. 2005, ApJS, {\bf 160}, 199
\bibitem{p06} Park, M., G., 2006, MNRAS, {\bf 367}, 1739
\bibitem{pss83} Pozdnyakov, A., Sobol, I. M., Sunyaev, R. A., 1983, Ap\&SS, {\bf 2}, 189
\bibitem{rcc06} Ryu, D. Chattopadhyay, I., Choi, E., ApJS, 2006, {\bf 166}, 410 [RCC06]
\bibitem{s57}Synge, J. L. 1957, {\it The Relativistic Gas} (Amsterdam: North Holland) [S57].
\bibitem{t48} Taub, A. H. 1948, Phys., Rev., {\bf 74}, 328.
\bibitem{w72} Weinberg, S., 1972, {\it Gravitation and Cosmology} (John WIlley and Sons, Inc).
\end{thebibliography}
\end{document}